\documentclass[aps,prb,twocolumn,superscriptaddress,showpacs]{revtex4-1}
\usepackage{graphicx}% Include figure files
\usepackage{color}
\usepackage{amsbsy}
\usepackage{amsmath}
\usepackage[normalem]{ulem}
\usepackage{lipsum}
\usepackage[rightcaption]{sidecap}
\usepackage{setspace}
\usepackage{placeins}
\begin{document}

\title{\fontfamily{phv}\selectfont{\Large{\bfseries{Jahn-Teller distortion driven magnetic polarons in magnetite}}}}
%\title{\fontfamily{phv}\selectfont{\Large{\bfseries{Magnetic polaron driven spin-orbital excitations in magnetite}}}}

\author{H. Y. Huang}
\affiliation{National Synchrotron Radiation Research Center, Hsinchu
30076, Taiwan}
\affiliation{Program of Science and Technology of Synchrotron Light Source, National Tsing Hua University, Hsinchu 30013, Taiwan}

\author{Z. Y. Chen}
\affiliation{Department of Physics, National Tsing
Hua University, Hsinchu 30013, Taiwan}

\author{R.-P. Wang}
\author{F. M. F. de Groot}
\affiliation{Department of Inorganic Chemistry and Catalysis, Utrecht University, Sorbonnelaan 16, 3584 CA Utrecht, The Netherlands}

\author{W. B. Wu}\author{J. Okamoto} \author{A. Chainani}
\affiliation{National Synchrotron Radiation Research Center, Hsinchu 30076, Taiwan}

\author{J.-S. Zhou}
\affiliation{Department of Mechanical Engineering, Texas Material Institute, University of Texas at Austin, Austin, Texas 78712, USA}

\author{H.-T. Jeng}

\affiliation{Department of Physics, National Tsing Hua University, Hsinchu 30013, Taiwan}

\author{G. Y. Guo}
\affiliation{Department of Physics, National Taiwan University, Taipei 10617, Taiwan}
\affiliation{Physics Division, National Center for Theoretical Sciences, Hsinchu 30013, Taiwan}

\author{Je-Geun Park}
\affiliation{Department of Physics and Astronomy, Seoul National University, Seoul 08826, Korea}
\affiliation{Center for Correlated Electron Systems, Institute for Basic Science, Seoul 08826, Korea}

\author{L. H. Tjeng}
\affiliation{Max Planck Institute for Chemical Physics of Solids, N$\ddot{o}$thnitzerstr. 40, 01187 Dresden, Germany} 
\author{C. T. Chen}
\affiliation{National Synchrotron Radiation Research Center, Hsinchu 30076, Taiwan}

\author{D. J. Huang}
\altaffiliation[Corresponding author:] {\emph{
djhuang@nsrrc.org.tw}} 
\affiliation{National Synchrotron Radiation Research Center, Hsinchu 30076, Taiwan} \affiliation{Department of Physics, National Tsing Hua University, Hsinchu 30013, Taiwan}

\date{\today}

\begin{abstract}
The first known magnetic mineral, magnetite (Fe$_3$O$_4$), has unusual properties which have fascinated mankind for centuries; it undergoes the Verwey transition at  $T_{\rm V}$~$\sim$120~K with an abrupt change in structure and electrical conductivity. The mechanism of  the Verwey transition however remains contentious. Here we use resonant inelastic X-ray scattering (RIXS) over a wide temperature range across the Verwey transition to identify and separate out the magnetic excitations derived from nominal Fe$^{2+}$ and Fe$^{3+}$ states.
Comparison of the RIXS results with crystal-field multiplet calculations shows that the spin-orbital $dd$ excitons of the Fe$^{2+}$ sites arise from a tetragonal Jahn-Teller active polaronic distortion of the Fe$^{2+}$O$_6$ octahedra. 
These low-energy excitations, which get weakened for temperatures above 350~K but persist at least up to 550~K, are distinct from optical excitations and best explained as magnetic polarons.

\end{abstract}
\maketitle

Since its first X-ray structural elucidation by W. H. Bragg a hundred years ago\cite{Bragg15} and the discovery of the Verwey transition \cite{Verwey39,Walz02}, Fe$_3$O$_4$ has received much attention for decades. Even today, it attracts significant scientific and technological interest for its applications in ultrafast magnetic sensors\cite{Jong13}, palaeomagnetism\cite{Almeida14,Jacob16}, nanomedicine carriers\cite{Veintmillas14}, etc. Fe$_3$O$_4$ becomes ferrimagnetic below T$_c$ $\sim$ 850 K, followed by an abrupt decrease in its electrical conductivity by two orders of magnitude as the temperature is cooled below  $T_{\rm V}$. Verwey first suggested a Fe$^{2+}$-Fe$^{3+}$ charge-ordering (CO) model as the driving force of this transition.  
There are two major schools of interpretation: the first one interprets the Verwey transition as a transition driven by charge/orbital ordering \cite{Wright01,Wright02,Jeng04,Leonov04,Huang06a,Jeng06,Nazarenko06,Schlappa08,Senn12a,Senn12b,Senn15}. The second one exploits the mechanism of a lattice distortion driven electron-phonon interaction which opens a gap at the Fermi energy when the temperature is lowered below the transition temperature $T_{\rm V}$ \cite{Garcia00,Piekarz06,Subias04,Chainani95,Park97,Taguchi15, Schrupp05}. %However, the change in the magnetic behavior of magnetite across the Verwey transition has not been well established.  

\begin{figure}
\includegraphics[width=0.95\columnwidth]{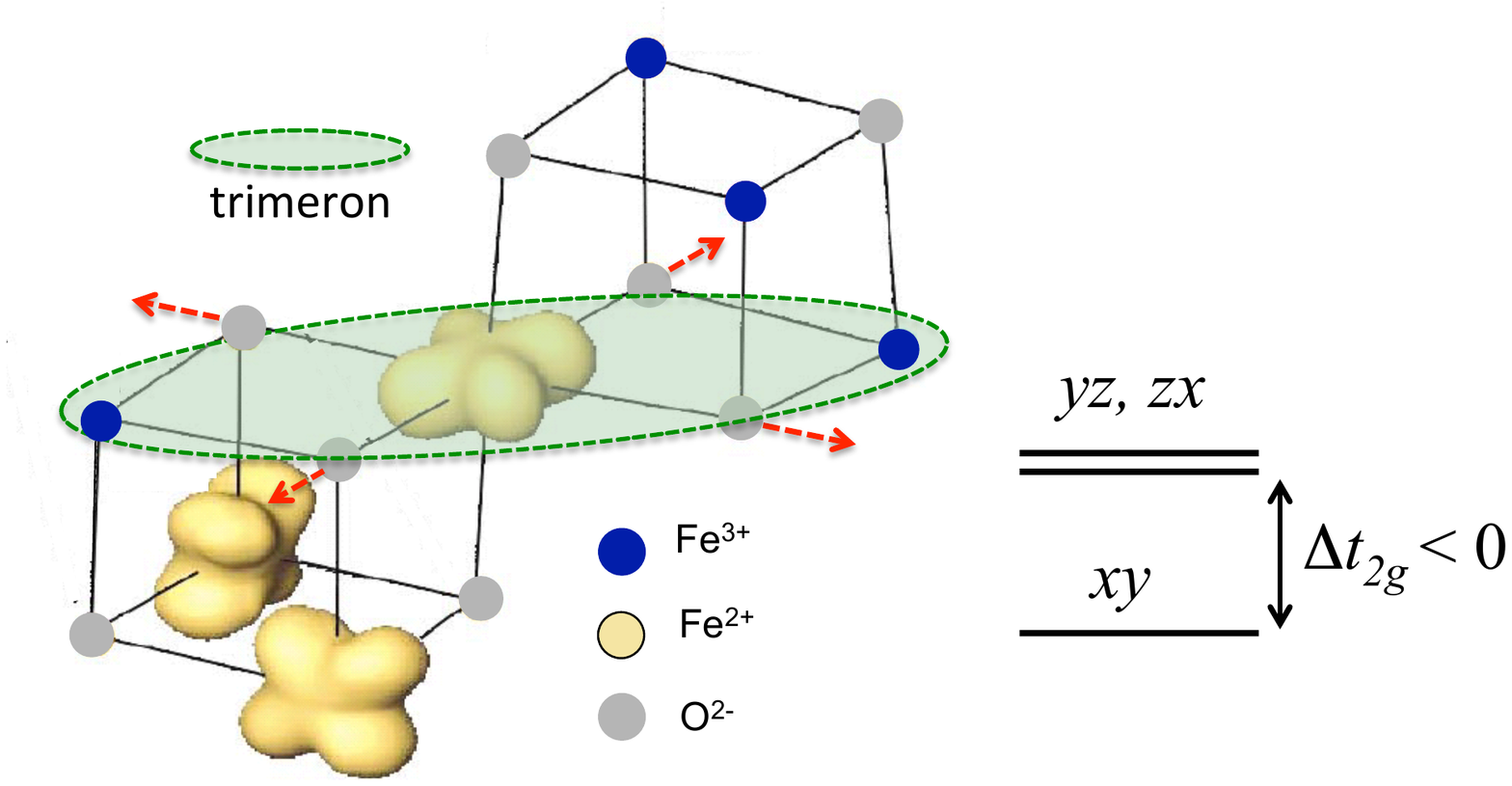}
\caption{{\bf The trimeron scenario and corresponding $t_{2g}$ energy-level splitting of Fe$_3$O$_4$.} Left: Illustration of the orbital ordering of B-site Fe$^{2+}$ and the corresponding trimeron. A trimeron is indicated with a dashed oval. The elongation of the four Fe-O bonds in the $xy$ plane are indicated with arrows. Right: The corresponding $t_{2g}$ energy-level splitting for a Fe$^{2+}$ ion in a negative $\Delta_{t_{2g}}$ crystal field.}\label{Fig1}
\end{figure}

\begin{center}
\begin{figure*}[ht!]
\includegraphics[width=2\columnwidth]{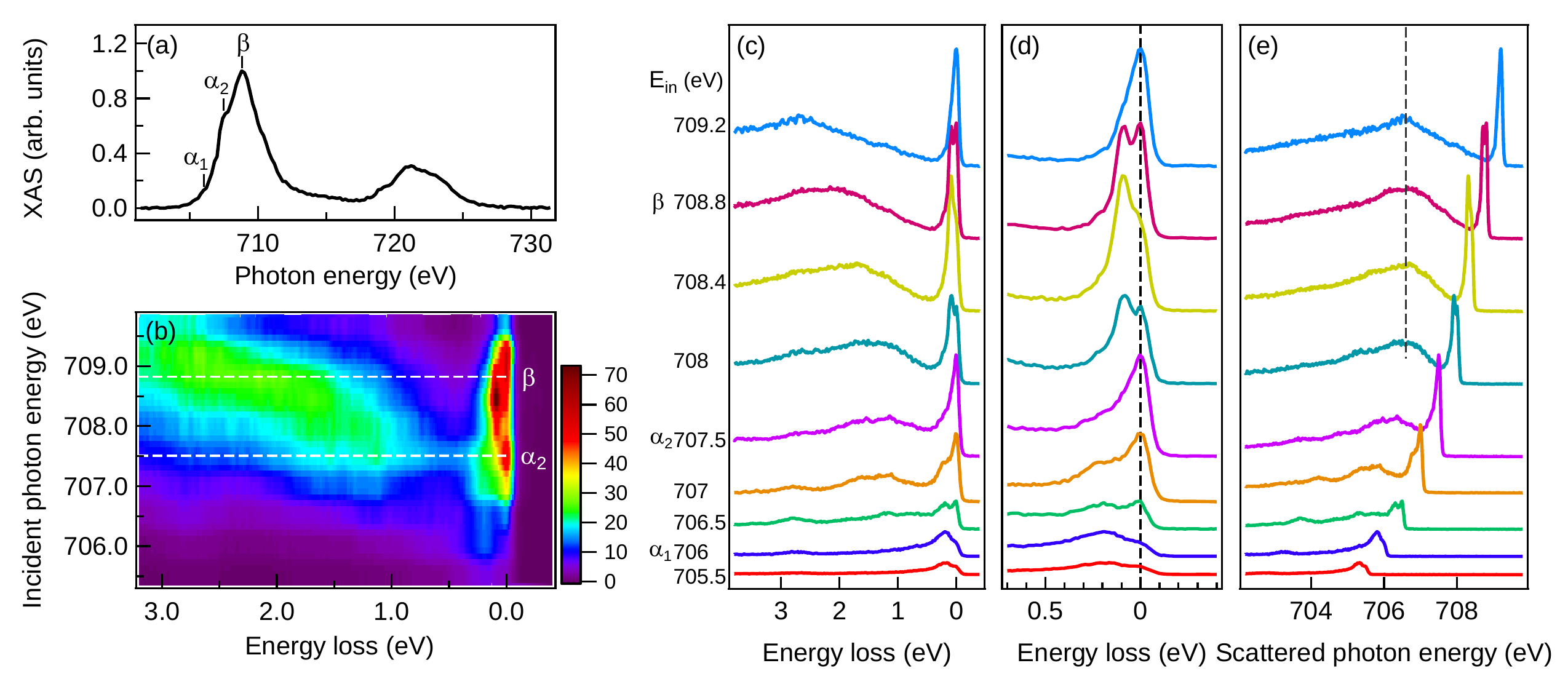}
\caption{{\bf RIXS measurements of Fe$_3$O$_4$.} (a) Fe $L$-edge XAS spectrum measured in the fluorescence yield mode through the summation of all inelastic X-ray intensities. The XAS is plotted with correction for self-absorption. The incident X-ray energy resolution was 0.5 eV.  $\alpha_1$,  $\alpha_2$ and $\beta$ denote the incident photon energies at 706~eV, 707.5~eV and 708.8~eV, respectively. (b) Color map of RIXS intensity  after correction for self-absorption in the plane of incident photon energy vs. energy loss. (c) \& (d) RIXS spectra plotted in terms of energy loss. (e) RIXS spectra plotted in terms of scattered X-ray energy. All RIXS spectra are after correction for self-absorption. They  were recorded by using $\pi$-polarized incident X-rays under the scattering geometry of the scattering angle $\phi = 90^{\circ}$ and the incident angle $\theta = 20^{\circ}$ at 80~K.}
\label{Fig2}
\end{figure*}
\end{center}

Although numerous investigations have been carried out to verify the charge localization on the octahedrally ($O_h$) coordinated B sites, the charge-ordering pattern of magnetite is subtle and still elusive \cite{Garcia00,Subias04}. While it is agreed that the  charge disproportionation involves changes in the nominal Fe$^{2+}$ and Fe$^{3+}$  states associated with the B-sites, recent X-ray diffraction studies of the low-temperature phase of magnetite microcrystals \cite{Senn12b,Senn15} revealed that the $t_{2g}$ electrons of the B-site are not fully localized in the form of Fe$^{2+}$ states. Instead, the electrons are distributed over linear three-Fe-site units termed ``trimerons," (consisting of one Fe$^{2+\delta}$ and two Fe$^{3-\delta}$ sites) which are coupled to the tetragonal ($T_{d}$) Jahn-Teller distortion of B-site Fe$^{2+}$O$_6$ octahedra, as illustrated in Fig.~\ref{Fig1}. To the first approximation, the B-site Fe$^{3+}$O$_6$ octahedra are Jahn-Teller inactive. 
The tetragonal distortion of B-site Fe$^{2+}$O$_6$ octahedra removes the degeneracy of $t_{2g}$ orbitals, in going from $O_h$ symmetry to $D_{4h}$ symmetry. In the absence of spin-orbit coupling, an effective energy separation $\Delta_{t_{2g}}$ between $d_{xy}$ and $d_{yz}/d_{zx}$ is created if the four Fe-O bonds in the $xy$ plane are elongated.  The trimeron scenario then indicates that the Verwey transition is essentially an ordering of trimerons. The authors further conclude that trimeron correlations might persist in the cubic phase at temperatures above $T_{\rm V}$, in line with the existence of the short-range order above $T_{\rm V}$ from results of neutron/X-ray diffuse scattering \cite{Bosak14}, X-ray absorption \cite{Subias05}, optical conductivity \cite{Park98}, photoemission \cite{Chainani95, Taguchi15} and anomalous phonon broadening \cite{Hoesch13}.

To the best of our knowledge, the relation of the local tetragonal distortion field of Fe$^{2+}$ ions with the magnetic excitations of magnetite has not been reported to date. The issue of the quenched/unquenched orbital moment at the Fe$^{2+}$ sites also remains controversial\cite{Huang04,Goering06}. Here we present measurements of resonant inelastic X-ray scattering (RIXS)\cite{Ament11}  at the Fe $L_3$-edge on magnetite to reveal the low-energy spin-orbital excitations of Fe$^{2+}$ ions in both the monoclinic and cubic phases. In combination with crystal-field multiplet calculations, we show the the existence of magnetic polarons in magnetite which is driven by Jahn-Teller distortion.

\vspace{3mm}

\noindent {\fontfamily{phv}\bfseries Results}
\vspace{3mm}

\begin{figure}[h]
\includegraphics[width=0.95\columnwidth]{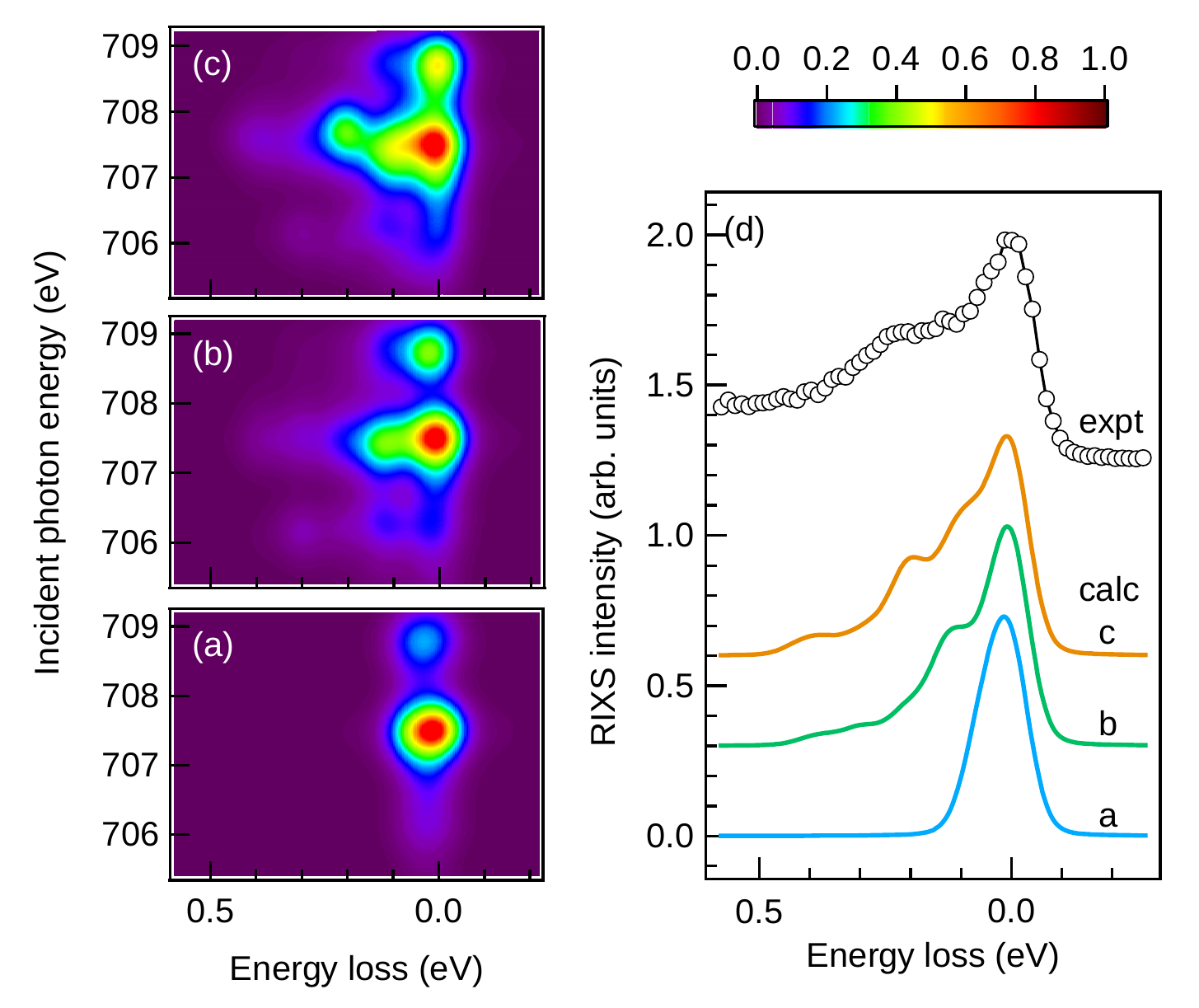}
\caption{{\bf Calculated RIXS of Fe$^{2+}$ in comparison with measurements.} Calculated RIXS intensity maps (a), (b) and (c) correspond to $H_{\rm ex} = 0, {\Delta_{t_{2g}}} = 0$;  $H_{\rm ex} = 90$~meV, ${\Delta_{t_{2g}}} = 0$, and $H_{\rm ex} = 90$~meV, ${\Delta_{t_{2g}}} = -24$~meV, respectively.
The core-hole lifetime width is set to 200 meV and the final-state lifetime width is set to 10 meV. These intensity maps present the average RIXS intensity for the magnetic easy axis along the [100], [010] and [001] directions and are plotted after Gaussian broadening of width 500 meV and 80 meV for the incident photon energy and the energy loss, respectively.
(d) Comparison of measured (expt) and calculated (calc) RIXS spectra. Open circles are measurements with incident X-rays of 707~eV; solid lines a, b, and c are the corresponding RIXS spectra extracted from panels (a), (b) and (c) with the incident X-ray which induces the maximum quasielastic peak.}\label{Fe2+}
\end{figure}

\noindent {\bf RIXS at the Fe $L_{3}$-edge.}  Figure \ref{Fig2}(a) shows the Fe $L$-edge absorption spectrum (XAS) of magnetite. By comparing with crystal-field multiplet calculations (See Supplementary Fig. 6.), it is understood
that the absorption-energy centroid of Fe$^{2+}$ ions is lower than that of Fe$^{3+}$ ions by $\sim 1-2$~eV, consistent with earlier work\cite{Kuiper97, Chen04, Arenholz06}. Accordingly, the features labelled  ``$\alpha_1$" and ``$\alpha_2$" at X-ray energies of 706.0 eV and 707.5~eV originate from the absorption of octahedrally coordinated B-site Fe$^{2+}$ states, while the maximum intensity feature ``$\beta$" is dominated by intensity from the 
the Fe$^{3+}$ ions of both the B-site octahedral and A-site tetrahedral symmetries. 

The color map of RIXS intensity in the plane of incident photon energy vs. energy loss shown in Fig. \ref{Fig2}(b) presents the evolution of the RIXS spectral profile associated with Fe$^{2+}$ and Fe$^{3+}$ ions as detailed in the following.  
When the incident X-ray energy set to below $\alpha_2$, we observed $dd$ excitations of Fe$^{2+}$ with energy loss at 2.8~eV, 1.65~eV and 1.16~eV shown in Fig. \ref{Fig2}(c), and also a broad excitation centered at 200~meV shown in Fig. \ref{Fig2}(d). 
%In addition, phonons are expected to contribute to the RIXS cross section for energy loss below 85~meV \cite{Gasparov09,Hoesch13,Verble74,Kumar14}, which cannot be resolved due to the finite instrument resolution of the present RIXS setup. 
If the incident X-ray energy goes beyond $\alpha_2$, the 1.16-eV $dd$ excitation of Fe$^{2+}$ begins to evolve into a fluorescence that has a constant X-ray emission energy independent of incident energy, as indicated by the dashed line in Fig. \ref{Fig2}(e). With the incident X-ray energy set to $\beta$, RIXS excitations arise mostly from Fe$^{3+}$ ions of octahedral or tetrahedral symmetry.

Figure \ref{Fig2}(d) shows  two RIXS features centered at 90 meV and 200 meV in a magnified plot of energy loss below 0.7~eV. Measurements carried out by varying the scattering angle suggested that these two low-energy excitations do not disperse in momentum space. (See Supplementary Fig. 3.) The  200-meV excitation has a full width at half maximum (FWHM) larger than the instrumental energy resolution. This broad RIXS feature resonates near the $L_3$-edge of Fe$^{2+}$ and almost disappears for incident energy above 708~eV, at which the other excitation centered at 90 meV emerges. The 90-meV excitation has a FWHM nearly equal to the instrumental energy resolution and resonates at 708.4 eV. The distinct  incident X-ray energies for these resonant excitations indicates that the 200-meV and 90-meV features arise from Fe$^{2+}$ and Fe$^{3+}$ states, respectively. 

\vspace{1mm}
\noindent {\bf Crystal-field multiplet calculations.}  To characterize the origin of the observed excitations centered at 200 meV, we undertook crystal-field multiplet calculations by using simulation programs CTM4RIXS \cite{CTM4RIXS} and {MISSING} \cite{MISSING}.
Figures \ref{Fe2+}(a), \ref{Fe2+}(b) and \ref{Fe2+}(c) show the low-energy RIXS excitations of Fe$^{2+}$ in the form of incident photon energy vs. energy loss maps, calculated using the same geometry as that of the experiments.

 The calculated spectra are obtained as an average of the spectra calculated for magnetic domains with the easy axis along the [100], [010] and [001] directions. The crystal field parameter $10Dq$ was set to 1.13~eV and the Slater integral was reduced to 79\% for accurately reproducing the $dd$ excitation energies. With only the spin-orbit coupling strength $\zeta_{3d}$~=~52~meV included, there exists low-energy excitation at 64~meV, but the 200~meV is not reproduced (Fig. \ref{Fe2+}(a)). If an effective exchange field $H_{\rm ex}=90$~meV is included without the tetragonal distortion, these excitations are split further with the excitation energy centroid at 132~meV, but still the 200~meV feature is not obtained (Fig. \ref{Fe2+}(b)). We, therefore, need to either increase the effective exchange field to nearly 200 meV, or include the effect of the tetragonal distortion of FeO$_6$ octahedra. It is, however, unreasonable to use an exchange field much larger than the spin wave energy or the exchange field of Fe$^{3+}$, 90~meV. Hence, we included a tetragonal distortion for calculating the RIXS spectrum of Fe$^{2+}$ in magnetite.
Figure \ref{Fe2+}(c) reveals that the calculated RIXS obtained on including the tetragonal distortion, exchange interaction and $3d$ spin-orbit coupling matches fairly well with the experimental data. The results further indicate that within an excitation energy of the first 400-meV, there are effectively 15 separate states from Fe$^{2+}$, as the $^{5}T_{2g}$ ground state is split by the combination of these interactions. Further improvements in energy resolution is expected to reveal these low-lying excitations in ever increasing detail. 

\vspace{3mm}

\noindent {\fontfamily{phv}\bfseries Discussion}

\vspace{3mm}

Because the observed RIXS excitations exist in the cross-polarization geometry of a 90$^\circ$ scattering (See Supplementary Fig. 3(a).), it rules out the orbital excitations of the same symmetry\cite{Veenendaal06}.  
The origin of the 90-meV and 200-meV excitations is thus different from those of  optical gap\cite{Park98}, photoemission gap\cite{Chainani95,Park97,Taguchi15, Schrupp05}  and the low-temperature activation energy obtained from electrical resistivity\cite{Kuipers76}.
Like magnetic excitations observed in the RIXS of cuprates\cite{Ament11,Jia14, Huang16} and nickelates \cite{deGroot98,Haverkort10}, the observed 90-meV excitation results from spin-flip excitations of Fe$^{3+}$ ions because its energy scale nearly corresponds to the energy of spin waves observed in inelastic neutron scattering \cite{McQueeney06, McQueeney07} and the RIXS cross section for phonons is much smaller than that of magnetic excitation.

For the broad 200-meV RIXS feature associated with the octahedral Fe$^{2+}$ states, the excitation energy is too large to be explained in terms of only spin-flip excitations. In order to understand the nature of this feature, we carefully checked the effect of the local Jahn-Teller distortion to explain the energy of the observed excitations. According to Hund's rule, out of the six $3d$ electrons of the Fe$^{2+}$ ion, five $3d$ electrons occupy spin-up states ($t_{2g}^{3\uparrow}$, $e_{g}^{2\uparrow}$); the remaining one electron occupies one of the three spin-down orbitals  ($t_{2g}^{1\downarrow}$). The ground state of the octahedral Fe$^{2+}$ ion is a high-spin $^{5}T_{2g}$ state with $S=2$.  When the spin-orbit effect of $3d$ electrons couples a pseudo-orbital angular momentum $\widetilde{L}=1$ to $S=2$, the  $^{5}T_{2g}$ state splits into three manifolds of pseudo-angular momenta $\widetilde{J}=1, 2$ ~and~3 with degeneracies 3, 5 and 7, respectively.

\begin{figure}
\includegraphics[width=0.95\columnwidth]{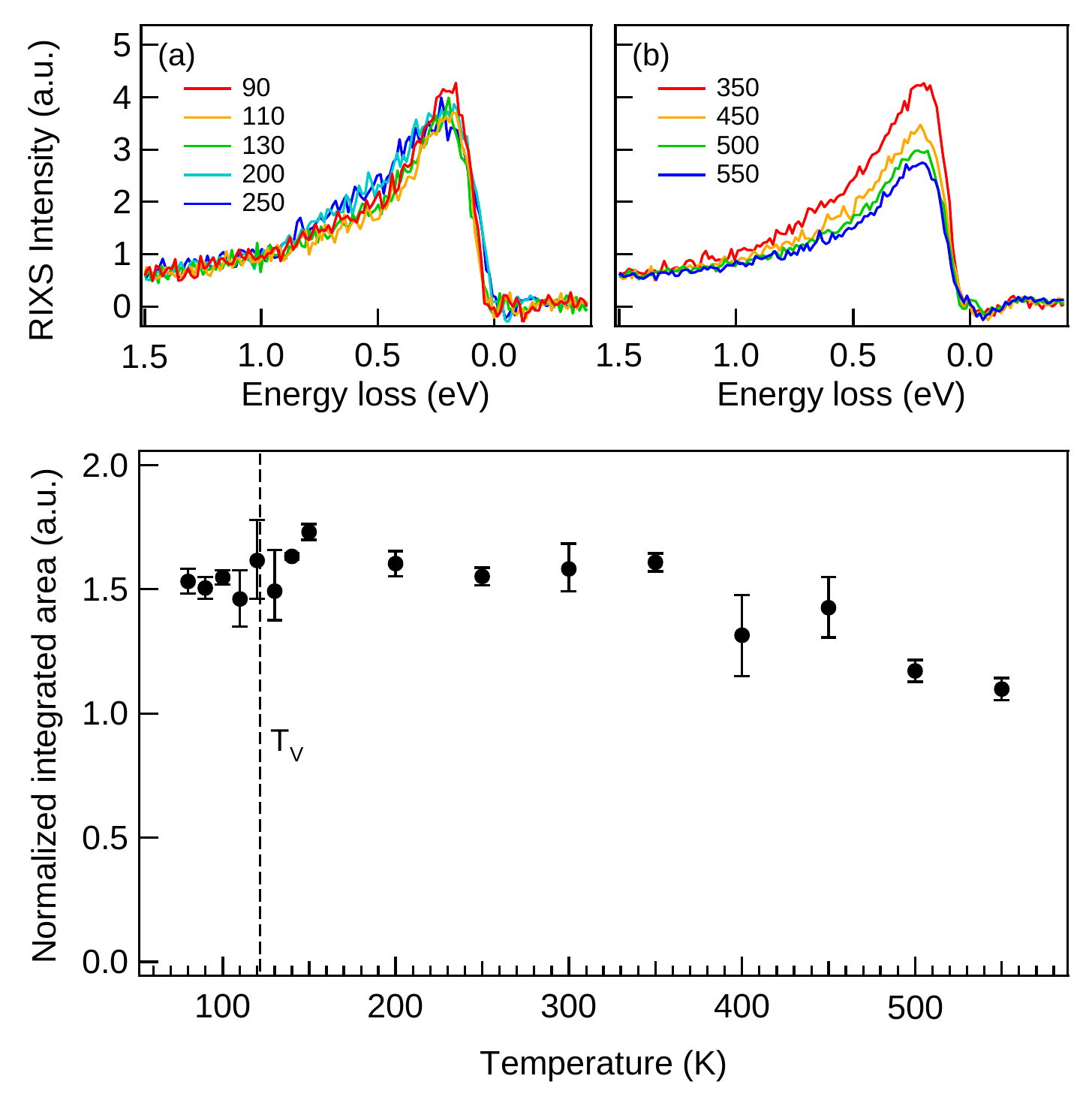}
\caption{{\bf Temperature-dependent 200-meVspin-orbital excitations of Fe$_3$O$_4$.} (a) \& (b) RIXS spectra after the subtraction of the elastic component at selected temperatures between 90~K and 550~K. The spectra were recorded with the incident X-ray energy set to 706~eV (c) Plot of the normalized integrated 200-meV RIXS intensity after the subtraction of the elastic component versus temperature. The RIXS data comprise four runs of experimental results.}\label{Temperature}
\end{figure}

From an extensive set of RIXS calculations of Fe$^{2+}$ with varied tetragonal distortions (See Supplementary Fig.~5.), we found that calculations using $H_{\rm ex}=90$~meV and $\Delta_{t_{2g}}$ about $-24$~meV  explain the measured RIXS spectra most satisfactorily. Figure \ref{Fe2+}(d) presents calculated RIXS of Fe$^{2+}$ in comparison with measurements of the incident X-ray energy set to 707~eV, at which the 200-meV RIXS feature is most pronounced. The negative value of $\Delta_{t_{2g}}$ signifies that the energy of $d_{xy}$ is lower than that of $d_{yz}/d_{zx}$, i.e., tetragonally distorted Fe$^{2+}$O$_6$ octahedra with elongated Fe-O bonds in the $xy$ plane. This shows that the tetragonal distortion is directly related to a polaronic distortion of the Fe$^{2+}$O$_6$ octahedra, which in turn couple to the 
neighbouring Fe$^{3+}$O$_6$ octahedra constituting the trimerons, although, as mentioned earlier, they are Jahn-Teller inactive in the first approximation.  In short, the observation of spin-orbital excitations driven by a tetragonal Jahn-Teller polaronic distortion provides evidence for magnetic polarons in magnetite. 

The magnitude of obtained $\Delta_{t_{2g}}$ is comparable with  the $3d$ spin-orbit coupling strength and thus confirms the observation of the unquenched  orbital moment \cite{Huang04}, which is known from work on Fe$^{2+}$ impurities in MgO thin films \cite{Haupricht10}. 
These results are also consistent with conclusions of band-structure calculations \cite{Jeng04, Leonov04,Jeng06} and  the recent X-ray diffraction study on magnetite\cite{Senn12a, Senn15}. Band-structure calculations using the monoclinic $P2/c$  crystal structure of magnetite \cite{Jeng04} also give the energy splitting  $\sim$~50 meV between minority-spin $d_{xy}$ and $d_{yz}/d_{zx}$ 
bands at the $\Gamma$ point, conforming to the deduced $\Delta_{t_{2g}}$.

Figure \ref{Temperature} plots the temperature-dependent RIXS spectra with the incident X-ray energy set to the pre-edge absorption at 706~eV which is denoted {$\alpha_1$}, an energy below $\alpha_2$ by 1.5~eV and at which the RIXS arises predominantly from octahedral Fe$^{2+}$ ions with a negligible contribution from Fe$^{3+}$ and the elastic component is weak. The results show that, when the temperature is varied across $T_{\rm V}$, the spin-orbital excitation of 200 meV does not abruptly change its intensity and persists at least up to 550 K, albeit with a gradual decrease above 350 K. We interpret this as a gradual weakening of the polarons.  
RIXS results shown here serve as a fast probe to snapshot the dynamic lattice-spin-orbital excitations of Fe$_3$O$_4$. 
These temperature-dependent RIXS results indicate that the FeO$_6$ octahedra are already locally distorted in the cubic phase of magnetite, with the existence of the short-range order above $T_{\rm V}$ \cite{Bosak14}. 
%Thus, magnetic polarons associated with the trimerons are responsible for the high conductivity in the cubic  phase  at $T >T_{\rm V}$, while the Verwey transition is due to ordering of these polarons. 

To summarize, our results demonstrate the usefulness of RIXS to unravel the local electronic structure of a mixed-valence compound by selecting the energy and polarization of incident  X-rays.  We revealed $dd$ excitons in magnetite that have an energy centroid 200~meV and arise from polaronic distortion driven spin-orbital excitations, which are best explained as magnetic polarons. We also applied crystal-field multiplet calculations to obtain the $t_{2g}$ crystal field ${\Delta_{t_{2g}}} = -24$~meV induced by the tetragonal Jahn-Teller distortion. These results are consistent with the trimeron mechanism for the Verwey transition. It would be interesting to carry out RIXS experiments with an improved energy resolution to study the change of spin-orbital excitations across the Verwey transition.

% and to measure the multiphonon excitations of the $A_{\rm 1g}$ mode for the study of the electron-phonon coupling in magnetite.

\vspace{3mm}
\noindent {\fontfamily{phv}\bfseries Methods}
\vspace{3mm}

\noindent {\bf Crystal growth.} Single-crystal growth of magnetite
was carried out in an infrared image furnace in high-purity
argon gas (99.999\% purity) atmosphere. Measurements of the temperature-dependent specific heat and resistivity of the synthesized magnetite crystal showed that it exhibits a sharp first order Verwey transition at  $T_{\rm V}=122$~K . The synthesized single crystal has a chemical composition of Fe$_{3(1-\delta)}$O$_{4}$ with $|\delta| \leq 0.00018$, indicative of a nearly ideal chemical stoichiometry. See Supplementary Figs. 1 and 2 for the sample characterization.

\vspace{1mm}
\noindent {\bf RIXS measurements.} Using the AGM-AGS spectrometer at beamline 05A1 of the National Synchrotron Radiation Research Center (NSRRC) in Taiwan, we measured RIXS on a single-crystal Fe$_{3}$O$_{4}(001)$ at incident photon energies set to specific energies about the $L_3$($2p_{3/2}\rightarrow 3d$) X-ray absorption edge of Fe.  Both the scattering angle $\phi$ defined as the angle between the incident and the scattered X-rays, and  the incident angle $\theta$ from the crystal $ab$ plane were variable. The polarization of the incident X-ray was switchable between $\pi$ and $\sigma$  polarizations, i.e. the polarization within and perpendicular to the scattering plane, respectively, and the polarization of scattered X-rays was not analyzed. The energy bandwidth of the incident X-rays was 500~meV and the total RIXS energy resolution was $\sim$ 80~meV because the energy compensation method was used to ensure a high-resolution measurement in the  energy loss scheme \cite{Lai14}. The beam diameter of incident X-ray at the sample is about 0.5~mm. See Supplementary Fig.~3(a) for the scattering geometry.

\vspace{1mm}
\noindent {\bf Crystal-field multiplet calculations\cite{Groot05}.}
 The starting point of the crystal field model is to approximate the transition metal Fe as an isolated atom surrounded
by a distribution of charges that mimic the solid around the transition metal. The crystal field Hamiltonian is regarded as a perturbation to the atomic Hamiltonian in terms of an electrostatic potential $\Phi({\mathbf r})$ that describes the surroundings. The atomic Hamiltonian include the spin-orbit coupling of $3d$ electrons and the electron-electron interaction is parameterized by the Slater-Condon parameters $F^{k}$ and $G^{k}$ through a reduction factor.  The Fe $L_3$-edge RIXS spectral intensity is calculated using the Kramers-Heisenberg formula 
\begin{equation} %\nonumber
I(\omega, \omega_{in})= \sum_{f} \left|  \sum_{m} \frac{\langle f |\hat{T}^{\dag}_{\boldsymbol{\epsilon}'} | m \rangle \langle m | \hat{T}_{\boldsymbol{\epsilon}} | 0 \rangle} {\omega_{in} +\mathit{E}_0 - H + \mathit{i}\Gamma}\right|^ 2 \delta (\omega + E_0 - E_{f}),
\end{equation} 
in which $H$ is the total Hamiltonian and $| 0\rangle$, $| m \rangle$ and $| f\rangle$ respectively represents the ground state of $n$ electrons in the $3d$ orbitals, intermediate states $| 3d^{n+1}2p^{5}\rangle$ in which one $2p$ core electron is promoted to the $3d$ orbital, and the final state of the $dd$ excitation, i.e. an excited state of $| 3d^{n}\rangle$. The corresponding energies of these three states are $E_0$, $E_m$, and $E_f$.  $\omega_{in}$ and ${\omega = \omega_{in} - \omega_{out}}$ are the incident photon energy and the energy loss, respectively.
The transition operator $\hat{T}_{\boldsymbol{\epsilon}}=  ({\boldsymbol{\epsilon}} \cdot \hat{r})\hat{a}^{\dag}_{3d}\hat{a}_{2p}$ (and h.c.) dictates the dipole transition process from Fe $2p$ to the $3d$ level (or from Fe $3d$ to $2p$), with the X-ray polarization $\boldsymbol{\epsilon}$ ($\boldsymbol{\epsilon}'$ for outgoing photon) either $\pi$ or $\sigma$; and 
$\Gamma$ is the inverse core-hole lifetime.


\begin{thebibliography}{00}

\bibitem{Bragg15} Bragg, W. H.  The Structure of Magnetite and the Spinels.
\emph{Nature} {\bf 95}, 561 (1915).


\bibitem{Verwey39} Verwey, E. J. W.  Electronic Conduction of Magnetite and its Transition Point at Low Temperatures. \emph{Nature} $\mathbf{144}$, 327 (1939).

\bibitem{Walz02} Walz, F. The Verwey transition--a topical review. \emph{J. Phys. Condens. Matter} $\mathbf{14}$, R285 (2002).

\bibitem{Jong13} De Jong, S. \emph{et al.} Speed limit of the insulator-metal transition in magnetite. \emph{Nature Mat.} 12, 882 (2013).

\bibitem{Almeida14} Almeida, T. P. \emph{et al.}  Visualized effect of oxidation on magnetic recording fidelity in pseudo-single-domain magnetite particles. \emph{Nat. Commun.} {\bf 5}, 5154 (2014).

\bibitem{Jacob16} Jacob, D. E. \emph{et al.} Redox-freezing and nucleation of diamond via magnetite formation in the Earth's mantle.
\emph{Nat. Commun.} {\bf 7}, 11891 (2016).

\bibitem{Veintmillas14} Veintemillas-Verdaguer, S. \emph{et al.} Magnetic nanocrsyatls for biomedical applications. \emph{Progress in Crystal growth and Characterization of Materials} {\bf 60}, 80 (2014).

\bibitem{Wright01} Wright, J. P., Attfield, J. P. \& Radaelli, P. G.  Long range charge ordering in magnetite below the Verwey transition. \emph{Phys. Rev. Lett.} \textbf{87}, 266401 (2001).

\bibitem{Wright02} Wright, J. P., Attfield, J. P. \& Radaelli, P. G. Charge ordered structure of magnetite below the Verwey transition. \emph{Phys. Rev. B} \textbf{66}, 214422 (2002).

\bibitem{Jeng04} Jeng,  H.-T.,  Guo, G. Y. \&  Huang, D. J. Charge-orbital ordering and Verwey transition in magnetite. \emph{Phys. Rev. Lett.} $\mathbf{93}$, 156403 (2004).

\bibitem{Leonov04} Leonov, I., Yaresko, A. N., Antonov, V. N., Korotin, M. A., \& Anisimov, V. I. (2004). Charge and Orbital Order in Fe$_3$O$_4$. \emph{Phys. Rev. Lett.} $\mathbf{93}$, 146404 (2004).
\bibitem{Jeng06} Jeng, H. T., Guo, G. Y., \& Huang, D. J. Charge-orbital ordering in low-temperature structures of magnetite: GGA+ U investigations. \emph{Phys. Rev. B} $\mathbf{74}$, 195115 (2006).

\bibitem{Huang06a} Huang, D. J. \emph{et al.} Charge-orbital ordering and Verwey transition in magnetite measured by resonant soft X-ray scattering.  \emph{Phys. Rev. Lett.} \textbf{96}, 096401 (2006).

\bibitem{Nazarenko06} Nazarenko, E., Lorenzo, J. E., Joly, Y., Hodeau, J. L., Mannix, D., \& Marin, C. (2006). Resonant X-ray diffraction studies on the charge ordering in magnetite. Phys. Rev. Lett. $\mathbf{97}$, 056403 (2006).

\bibitem{Schlappa08} Schlappa, J.  \emph{et al.} Direct observation of $t_{2g}$ orbital ordering in magnetite. \emph{Phys. Rev. Lett.} \textbf{100}, 026406 (2008).

\bibitem{Senn12b} Senn, M. S., Wright, J. P., \& Attfield, J. P. (2012). Charge order and three-site distortions in the Verwey structure of magnetite.  \emph{Nature} $\mathbf{481}$, 173 (2012).

\bibitem{Senn12a} Senn, M. S., Loa, I., Wright, J. P., \& Attfield, J. P. Electronic orders in the Verwey structure of magnetite. \emph{Phys. Rev. B} $\mathbf{85}$, 125119 (2012).


\bibitem{Senn15}  Senn, M. S., Wright, J. P., Cumby, J., \& Attfield, J. P. (2015). Charge localization in the Verwey structure of magnetite. \emph{Phys. Rev. B} $\mathbf{92}$, 024104 (2015).

\bibitem{Garcia00} J. Garc\'{\i}a, J.  \emph{et al.} Resonant "forbidden" reflections in magnetite.  \emph{Phys. Rev. Lett.} \textbf{85}, 578 (2000).

\bibitem{Subias04} Sub\'{i}as, G. \emph{et al.} Magnetite, a model system for mixed-valence oxides, does not show charge ordering. \emph{Phys. Rev. Lett.} $\mathbf{93}$, 156408 (2004).

\bibitem{Piekarz06} Piekarz, P., Parlinski, K., \& Ole{\'s}, A. M. Mechanism of the Verwey transition in magnetite. \emph{Phys. Rev. Lett.} \textbf{97}, 156402 (2006).

\bibitem{Chainani95} Chainani, A., Yokoya, T., Morimoto, T., Takahashi, T., \& Todo, S. High-resolution photoemission spectroscopy of the Verwey transition in Fe$_3$O$_4$. \emph{Phys. Rev. B} $\mathbf{51}$, 17976 (1995).

\bibitem{Park97} Park, J. H., Tjeng, L. H., Allen, J. W., Metcalf, P., \& Chen, C. T. Single-particle gap above the Verwey transition in Fe$_3$O$_4$.  \emph{Phys. Rev. B} $\mathbf{55}$, 12813 (1997).

\bibitem{Schrupp05} Schrupp, D. \emph{et al.} High-energy photoemission on Fe$_3$O$_4$: Small polaron physics and the Verwey transition. Europhys. Lett. {\bf70}, 789.(2005)

\bibitem{Taguchi15} Taguchi, M. \emph{et al.}  Temperature Dependence of Magnetically Active Charge Excitations in Magnetite across the Verwey Transition. \emph{Phys. Rev. Lett.} \textbf{115}, 256405 (2015).

\bibitem{Bosak14} Bosak, A.  \emph{et al.}  Short-range correlations in magnetite above the Verwey temperature. Phys. Rev. X $\mathbf{4}$, 011040 (2014).

\bibitem{Subias05} Sub\'{i}as, G.  J. Garc\'{i}a, J. \& J. Blasco, J.  EXAFS spectroscopic analysis of the Verwey transition in Fe$_3$O$_4$. \emph{Phys. Rev. B} $\mathbf{71}$, 155103 (2005).

\bibitem{Park98} Park, S. K., Ishikawa, T., \& Tokura, Y. Charge-gap formation upon the Verwey transition in Fe$_3$O$_4$. \emph{Phys. Rev. B} $\mathbf{58}$, 3717 (1998).


\bibitem{Hoesch13} M. Hoesch, P. Piekarz, A. Bosak, M. Le Tacon, M. Krisch, A. Kozlowski, A. M. Ole\'{s}, and K. Parlinski, \emph{Phys. Rev. Lett.} $\mathbf{110}$, 207204 (2013).

\bibitem{Huang04} Huang, D. J. \emph{et al.} Spin and Orbital Magnetic Moments of Fe$_3$O$_4$. \emph{Phys. Rev. Lett.} $\mathbf{93}$, 077204 (2004).

\bibitem{Goering06} Goering, E., Gold, S., Lafkioti, M., \& Sch\"{u}tz, G. Vanishing Fe 3d orbital moments in single-crystalline magnetite. Europhys. Lett. $\mathbf{73}$, 97 (2006).

%\bibitem{Goering06} E. Goering, M. Lafkioti, and S. Gold, \emph{Phys. Rev. Lett.} $\mathbf{96}$, 039701 (2006). 

%\bibitem{Huang06} D. J. Huang, H.-J. Lin, and C. T. Chen, \emph{Phys. Rev. Lett.} $\mathbf{96}$, 039702 (2006). 

\bibitem{Ament11} Ament, L. J. P., van Veenendaal, M., Devereaux, T. P., Hill, J. P. \& van den Brink, J. Resonant inelastic x-ray scattering studies of elementary excitations. \emph{Rev. Mod. Phys.} {\bf 83}, 705 (2011).

\bibitem{Lai14}  Lai, C. H. \emph{et al.} Highly efficient beamline and spectrometer for inelastic soft X-ray scattering at high resolution.  J. Synchrotron Radiat. $\mathbf{21}$, 325 (2014).

\bibitem{Kuiper97} Kuiper, P., Searle, B. G., Duda, L. C., Wolf, R. M., \& Van der Zaag, P. J. (1997). Fe $L_{2, 3}$ linear and circular magnetic dichroism of Fe$_3$O$_4$. J. Electron Spectrosc. Relat. Phenom. $\mathbf{86}$, 107 (1997).

\bibitem{Chen04} Chen, J., Huang, D. J., Tanaka, A., Chang, C. F., Chung, S. C., Wu, W. B., \& Chen, C. T. (2004). Magnetic circular dichroism in Fe $2p$ resonant photoemission of magnetite. \emph{Phys. Rev. B} $\mathbf{69}$, 085107 (2004).

\bibitem{Arenholz06} Arenholz, E., van der Laan, G., Chopdekar, R. V., \& Suzuki, Y. Anisotropic X-ray magnetic linear dichroism at the Fe $L_{2, 3}$ edges in Fe$_3$O$_4$. \emph{Phys. Rev. B} $\mathbf{74}$, 094407 (2006).

\bibitem{Verble74}  Verble, J. L. Temperature-dependent light-scattering studies of the Verwey transition and electronic disorder in magnetite.  \emph{Phys. Rev. B} $\mathbf{9}$, 5236 (1974).


\bibitem{Gasparov09} Gasparov, L. V., Rush, A., Guntherodt, G., \& Berger, H. Electronic Raman scattering in magnetite: Spin versus charge gap. \emph{Phys. Rev. B} $\mathbf{79}$, 144303 (2009).


\bibitem{Kumar14}  Kumar, A., Chaudhary, S., Pandya, D. K., \& Sharma, S. K. Evidence of electron-phonon and spin-phonon couplings at the Verwey transition in Fe$_3$O$_4$.  \emph{Phys. Rev. B} $\mathbf{90}$, 024302 (2014).

\bibitem{Jia14} Jia, C. J. \emph{et al.} Persistent spin excitations in doped antiferromagnets revealed by resonant inelastic light scattering. \emph{Nat. Commun.} {\bf 5}, 3314 (2014).

\bibitem{Huang16}	Huang, H. Y. \emph{et al.} Raman and fluorescence characteristics of resonant inelastic X-ray scattering from doped superconducting cuprates. \emph{Sci. Rep.} $\mathbf{6}$, 19657 (2016).


\bibitem{Veenendaal06} van Veenendaal, M. Polarization dependence of L-and M-edge resonant inelastic X-ray scattering in transition-metal compounds. \emph{Phys. Rev. Lett.} {\bf96}, 117404 (2006).


\bibitem{Kuipers76} Kuipers, A. J. M., \& Brabers, V. A. M. Thermoelectric properties of magnetite at the Verwey transition.  \emph{Phys. Rev. B} {\bf14}, 1401 (1976).

\bibitem{deGroot98} de Groot, F. M. F., Kuiper, P., \& Sawatzky, G. A. Local spin-flip spectral distribution obtained by resonant x-ray Raman scattering. \emph{Phys. Rev. B} {\bf57}, 14584 (1998).

\bibitem{Haverkort10} Haverkort, M. W. Theory of resonant inelastic X-ray scattering by collective
magnetic excitations. \emph{Phys. Rev. Lett.} {\bf 105}, 167404 (2010).

\bibitem{McQueeney06}McQueeney, R. J., Yethiraj, M., Montfrooij, W., Gardner, J. S., Metcalf, P., \& Honig, J. M. Investigation of the presence of charge order in magnetite by measurement of the spin wave spectrum.  \emph{Phys. Rev. B} $\mathbf{73}$, 174409 (2006).

\bibitem{McQueeney07} McQueeney, R. J., Yethiraj, M., Chang, S., Montfrooij, W., Perring, T. G., Honig, J. M., \& Metcalf, P. Zener double exchange from local valence fluctuations in magnetite. \emph{Phys. Rev. Lett.} $\mathbf{99}$, 246401 (2007).

\bibitem{CTM4RIXS} Stavitski, E., \& de Groot, F. M. F. The CTM4XAS program for EELS and XAS spectral shape analysis of transition metal L edges. \emph{Micron} $\mathbf{41}$, 687  (2010).

\bibitem{MISSING}  Claudia Dallera and Riccardo Gusmeroli, http://www.esrf.eu/computing/scientic/MISSING/.

\bibitem{Haupricht10} Haupricht, T. \emph{et al.} Local electronic structure of Fe$^{2+}$ impurities in MgO thin films: Temperature-dependent soft X-ray absorption spectroscopy study  \emph{Phys. Rev. B} $\mathbf{82}$, 035120 (2010).



\bibitem{Groot05} de Groot, F. M. F.  Multiplet effects in X-ray spectroscopy. \emph{Coord. Chem. Rev.}  $\mathbf{249}$, 31 (2005).
\end{thebibliography}
\end{document}